\documentclass{aastex}
\usepackage{emulateapj5}
\usepackage{apjfonts,pstricks,amsmath,epsfig,pst-node}

\makeatletter

\newenvironment{inlinefigure}{%
\def\@captype{figure}%
\noindent\begin{minipage}{0.999\linewidth}\begin{center}}
{\end{center}\end{minipage}\smallskip}
\makeatother

\newcommand\chandra{{\it Chandra\/ }}
\newcommand\hubble{{\it Hubble\/ }}
\newcommand\xmm{{\it XMM-Newton\/ }}

\slugcomment{Accepted to {\it The Astrophysical Journal Letters} }
\shorttitle{Photometric Redshifts for X-ray Selected Sources}
\shortauthors{Gonzalez \& Maccarone}
\begin{document}

\title{A Test of Photometric Redshifts for X-ray Selected Sources}
\author{ Anthony H. Gonzalez}
\affil{Department of Astronomy, University of Florida, Gainesville, FL 32611}
\author{Thomas J. Maccarone}
\affil{Scuola Internationale Superiore di Studi Avanzati, via Beirut, n. 2-4,
Trieste, Italy, 34014}
\begin{abstract}
  
  We test the effectiveness of photometric redshifts based upon galaxy
  spectral template fitting for X-ray luminous objects, using a sample
  of 65 sources detected by \chandra in the field of the
  Caltech Faint Galaxy Redshift Survey (CFGRS).  We find that sources
  with quasar-dominated spectra (for which galaxy spectral templates
  are not appropriate) are easily identified, and that photometric
  redshifts are robust for the rest of the sources in our sample.
  Specifically, for the 59 sources that are not quasar-dominated at
  optical wavelengths, we find that the photometric redshift estimates
  have scatter comparable to the field galaxy population in this
  region.  There is no evidence for a trend of increasing dispersion
  with X-ray luminosity over the range $L_X=10^{39} - 5\times10^{43}$
  erg s$^{-1}$, nor is there a trend with the ratio of X-ray to
  optical flux, $f_X/f_R$. The practical implication of this work is
  that photometric redshifts should be robust for the majority
  ($\sim$90\%) of the X-ray sources down to $f_X\approx10^{-16}$ erg
  s$^{-1}$ cm$^{-2}$ that have optical counterparts brighter than
  $R\approx24$. Furthermore, the same photometry can be easily used to
  identify the sources for which the photometric redshifts are likely
  to fail.  Photometric redshift estimation can thus be utilized as an
  efficient tool in analyzing the statistical properties of upcoming
  large \chandra and \xmm data sets and identifying interesting
  subsamples for further study.

\end{abstract}

\section{Introduction}  
\label{sec:introduction}
Photometric redshift estimation is a powerful tool in extragalactic
astronomy, and significant effort has been expended in recent years
developing robust redshift estimation techniques. The most widely
employed approach is to use spectral templates (either empirical or
from stellar synthesis models) to derive optimal fits to the observed
galaxy colors \citep[e.g.][]{lanzetta1996, fernandezsoto1999,
benitez2000,furusawa2000}.  This approach has been highly successful
for normal galaxies, achieving results as good as $\sigma_z=0.06(1+z)$
for the {\it Hubble} Deep Field \citep[HDF;][]{fernandezsoto1999,
benitez2000, furusawa2000}.  Extending upon this work,
\citet{budavari2001} and \citet{richards2001} have recently found that
photometric redshifts accurate to within $\Delta z=0.2$ can be
obtained for 70\% of quasars in the Sloan Digital Sky Survey -- even
at $z<2.2$, where quasar spectra lack a strong continuum break in the
Sloan $ugriz$ filters.

In this work we focus on X-ray luminous objects, asking whether a
self-consistent technique can be devised to obtain photometric
redshifts for X-ray selected samples.  Recent work on deep \chandra
fields indicates that the resolved X-ray background is comprised of a
variety of objects, including early-type ellipticals, starburst
galaxies, obscured active galactic nuclei (AGN), narrow- and
broad-line AGN, and quasars \citep[see
e.g.][]{mush2000,barger2001,horn2001,tozzi2001,stern2002,barger2002}.
Given the disparate nature of these sources, the effectiveness of
photometric redshifts is {\it a priori} unclear. Recent spectroscopic
observations in the \chandra Deep Field-North (CDF-N) highlight this
concern.  \citet{barger2002} find that, out of a sample of 182 hard
sources (2-8 keV detections) with spectroscopic follow-up,
approximately half show signatures of AGN activity in their spectra.

As argued by \citet{barger2002}, one can reasonably expect that
traditional galaxy spectral template fitting should be reliable for
the 50\% of sources that lack an AGN spectral signature.  Similarly,
it is likely that quasar spectral templates can be utilized for
AGN-dominated sources (although this expectation has not yet been
verified). It is unclear though whether either approach is robust for
the intermediate sources in which both the AGN and the galaxy stellar
population contribute significantly to the optical spectrum. Do
photometric redshifts based upon galaxy spectral template fitting
gradually degrade with increasing fractional luminosity contribution
from the AGN, or do they remain robust until the AGN contribution
reaches some critical level? The \hubble Deep Field, which is the
canonical field for testing photometric redshifts, offers little
insight. There are only six X-ray luminous sources with spectroscopic
redshifts in the HDF -- the majority of which are low-luminosity.
Photometric redshift comparisons in the HDF, such as the blind check
of \citet{cohen2000}, thus cannot test the reliability of photometric
redshift estimators in the interesting regime of properties.\footnote{
  One of these sources is an outlier in the blind photometric redshift
  tests of \citet{cohen2000}; however, it has also been argued by
  \citet{fernandezsoto2001} that the spectroscopic redshift for this
  source is incorrect. Two of the other objects are listed as faint
  X-ray sources in \citet{horn2001}, but do not appear in the
  $\approx$1 Ms \chandra catalogs of \citet{brandt2001}.}

There are strong practical motivations for using an expanded sample of
X-ray selected sources to assess the reliability of photometric
redshifts for these objects.  Foremost, an efficient means of redshift
estimation is required to maximize the return from large area \chandra
and \xmm surveys, such as the Lockman Hole survey ({\it Chandra}, PI:
Barger), the \chandra Multiwavelength Project
\citep[ChaMP][]{wilkes2000}, the {\it XMM-}LSS survey
\citep{pierre2001}, and the upcoming \chandra survey in the NOAO
Deep-Wide Field (9 sq. degrees, PIs: Jones \& Murray). A number of the
issues that these surveys aim to address, such as evolution in the AGN
X-ray luminosity 

\begin{inlinefigure}
\plotone{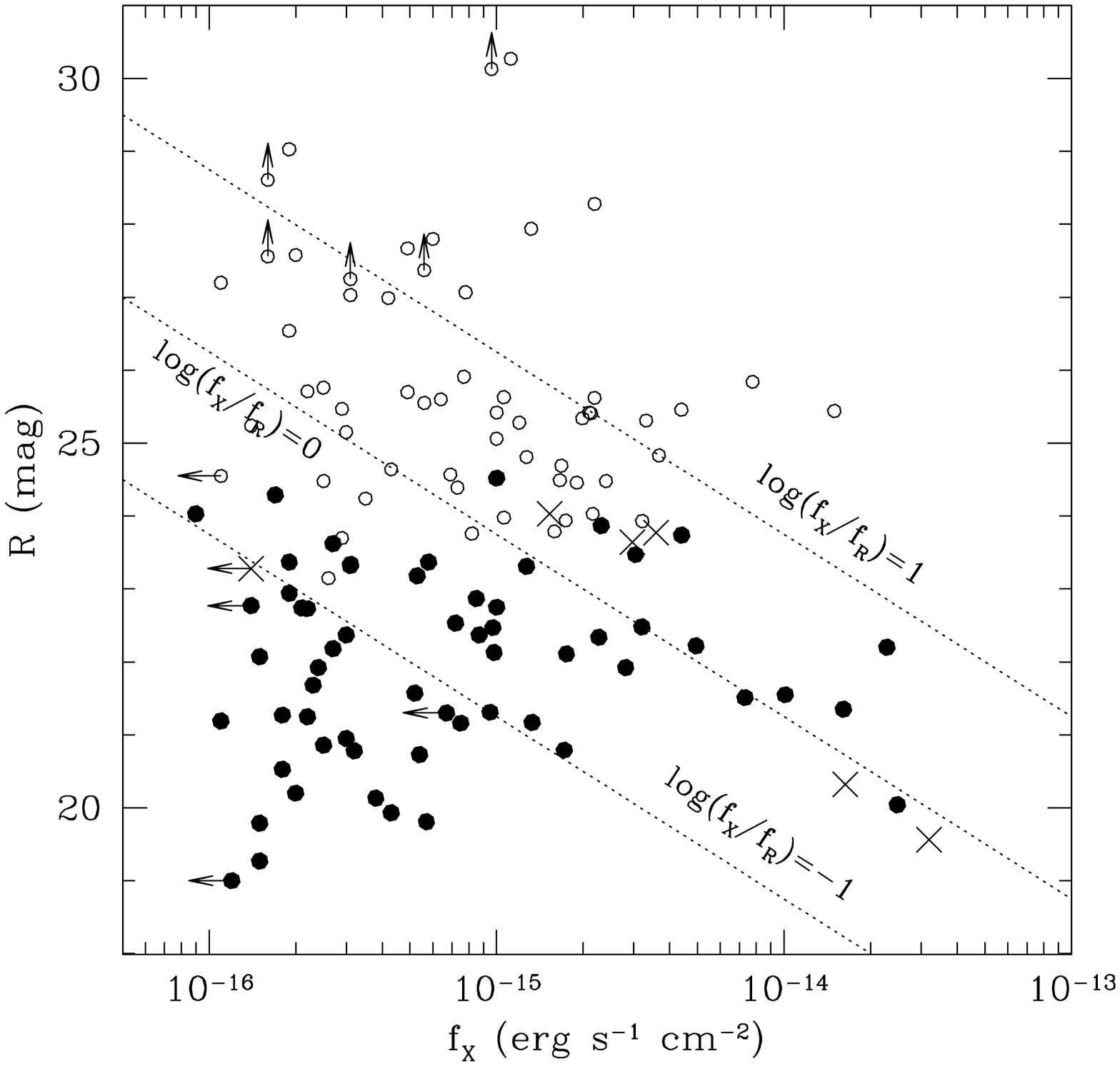} 
\figcaption{
    Optical $R$-band magnitude from \citet{barger2002} plotted against
  the full band 0.5-8 keV X-ray flux from \citet{brandt2001}.  The
  crosses correspond to the seven sources in our sample with colors
  that are best fit by a quasar spectral template.  The filled circles
  correspond to the other sources in our sample, while the open
  circles denote the rest of the extragalactic sources in the
  \citeauthor{barger2002} catalog that lie within the CFGRS region.
  Arrows denote sources that only have lower (upper) limits on their
  $R$-band magnitudes (X-ray fluxes). The dashed lines denote lines of
  constant $f_X/f_R$, as given by Equation 3 in \citet{horn2001}.
\label{fig:scatter}}
\end{inlinefigure}

\noindent function and evolution in the correlation function of
faint X-ray sources, are statistical in nature and do not require the
accuracy of spectroscopic redshifts.  For example, the survey in the
NOAO Deep Wide Field aims to study the X-ray spatial correlation
function using an expected $\sim$2200 sources (C. Jones 2002, private
communication). This project has no associated large spectroscopic
program and will depend upon the use of photometric redshifts to
achieve its key scientific goal.  If one can attain a level of
accuracy for these photometric redshifts comparable to what is
achieved for ``normal'' galaxies, photometric redshifts will be a
valuable tool. If not, then the limitations of photometric redshift
techniques must be established.

In this paper we assess the validity of photometric redshifts in the
Caltech Faint Galaxy Redshift Survey field
\citep[CFGRS,][]{cohen2000,hogg2000} using the \citet{brandt2001}
catalog of X-ray sources in the CDF-N. We first compare the scatter in
photometric redshift estimates for X-ray selected sources with a
control sample of quiescent galaxies, and quantify the fraction of
X-ray sources with bright optical counterparts for which traditional
galaxy spectral template fitting is valid. Next, we search for trends
between the redshift residuals and other physical quantities, testing
whether there is a range in X-ray luminosity ($L_X$) or flux ratio
($f_X/f_R$) over which photometric redshifts degrade. Finally, we ask
whether the objects for which photometric redshifts are likely to fail
(e.g. quasars) can be easily identified using the optical photometry.
The data are described in \S \ref{sec:data} and the results are
presented in \S \ref{sec:photoz}.  In \S \ref{sec:summary} we
summarize our work and discuss prospects for future studies 
with larger and fainter samples. We assume $\Omega_0=0.3$,
$\Omega_\Lambda=0.7$, and $H_0=100 h$ km s$^{-1}$ Mpc$^{-1}$.

\begin{inlinefigure}
\plotone{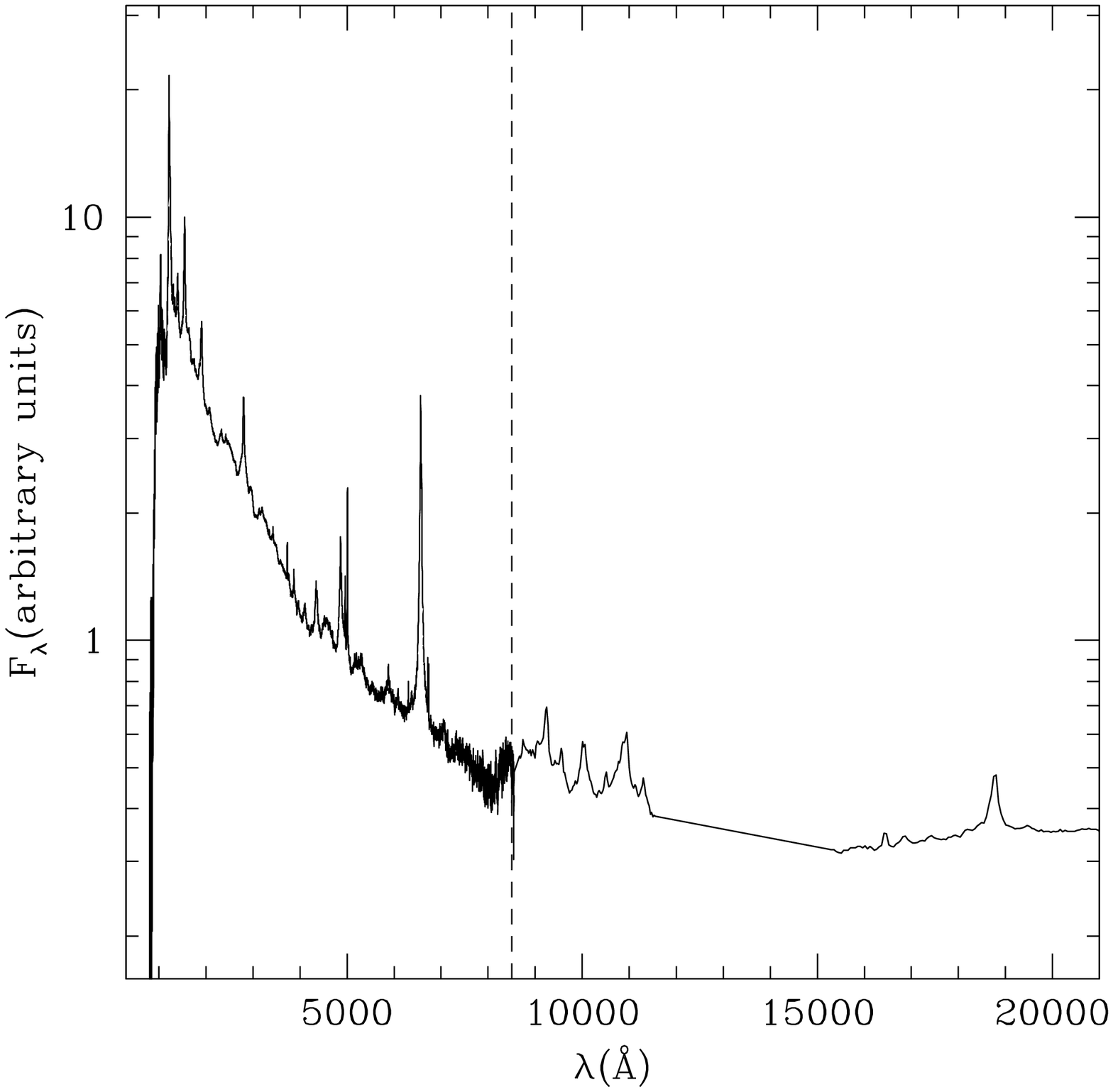} 
\figcaption{
    Template quasar spectra used in the photometric redshift code. The
  dashed line separates the SDSS composite spectra (short wavelength),
  which is corrected for stellar contamination, from the PDS 456
  spectra.  \citet{simpson1999} obtained $J-$ and $K-$band spectra for
  PDS 456; between these two bands we interpolate.
\label{fig:qsotemplate}}
\end{inlinefigure}

\section{Data}
\label{sec:data}

We focus on the CFGRS field due to the multiwavelength observations
and extensive spectroscopy available for this region. We include in
our sample X-ray sources detected by \citet{brandt2001} that lie
within the CFGRS, have published spectroscopic redshifts, and have
unambiguous optical counterparts.  The \citet{brandt2001} source list
is derived from the 975.3 ks \chandra image of the \chandra Deep Field
North (CDF-N), with minimum detectable fluxes near the aim point
corresponding to $f_x\approx1.9\times10^{-16}$ ergs s$^{-1}$ cm$^{-2}$
in the hard band (2-8 keV) and $f_x\approx2.9\times10^{-17}$ ergs
s$^{-1}$ cm$^{-2}$ in the soft band (0.5-2 keV).  Spectroscopic
redshifts \citep[]{cohen2000,horn2001,barger2002} and $U_n$, $G$,
${\cal R}$, $K_s$ photometry \citep{hogg2000} exist for 66 of these
sources, which are listed in Table \ref{tab:tab1}. For the majority of
these objects the spectra are also published in Figure 6 of
\citet{barger2002}.

\begin{figure*}
\plotone{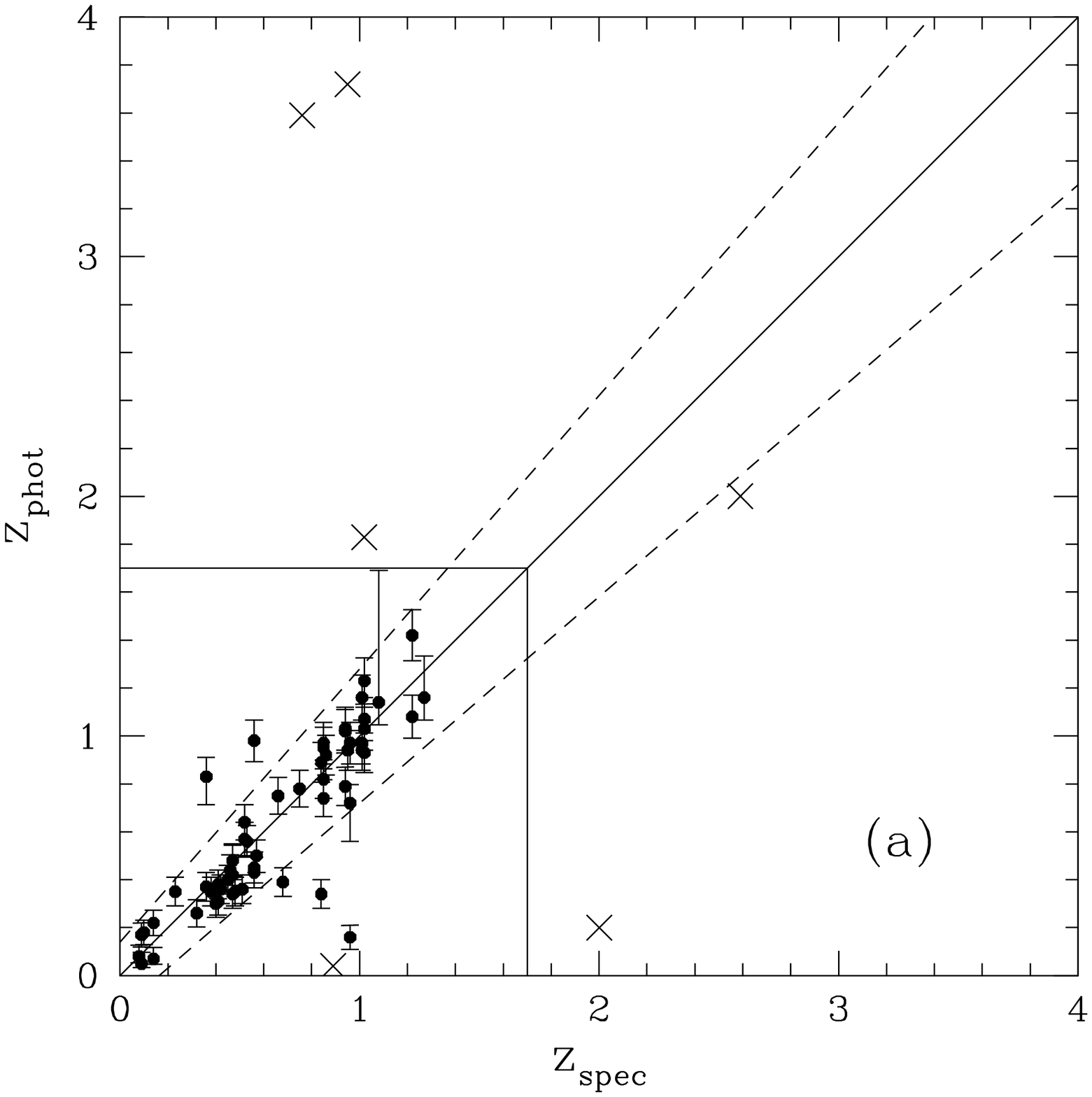}\plotone{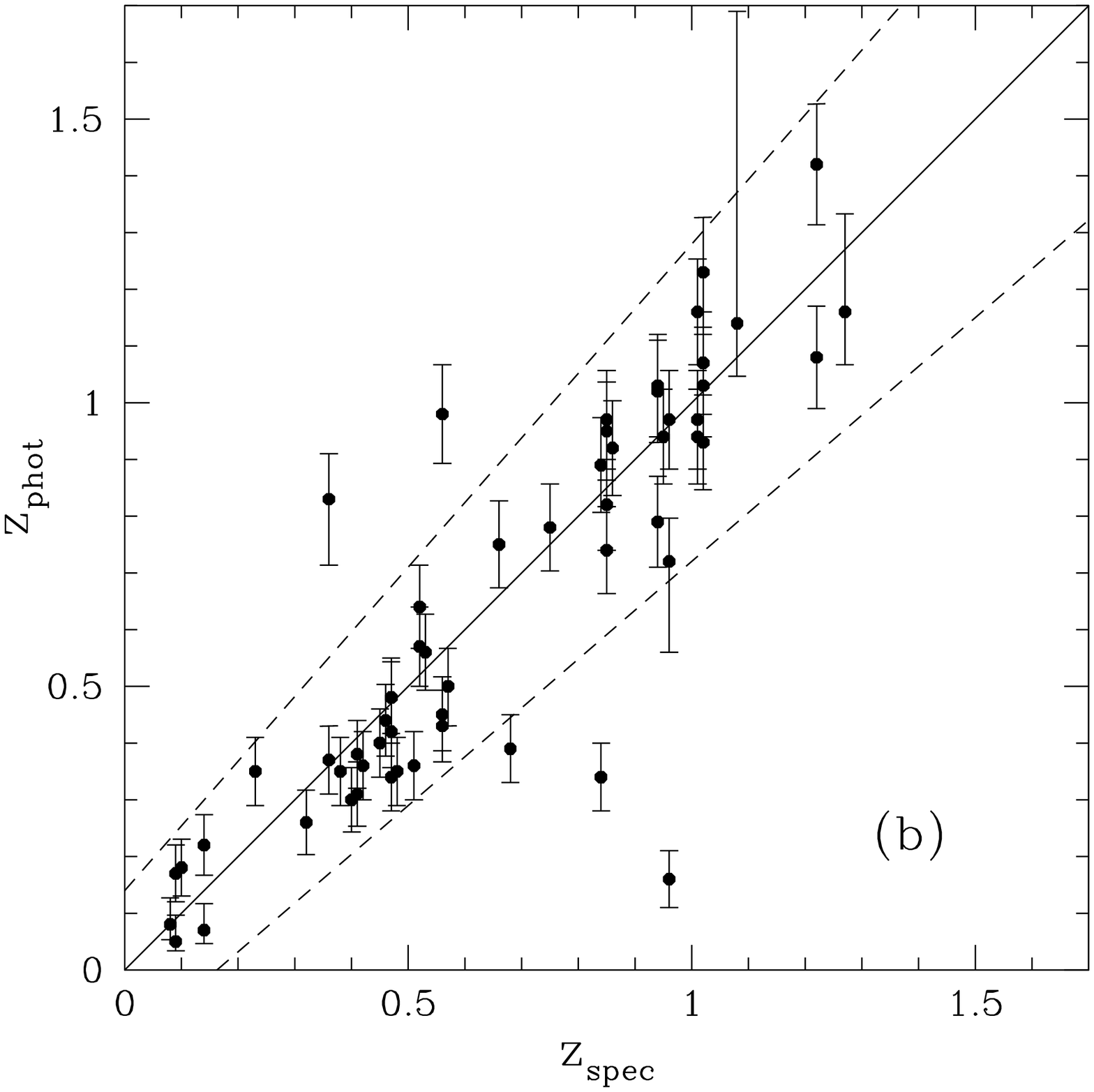}
\figcaption{
    ($a$) Photometric redshift vs. spectroscopic redshift. The crosses
  are the sources that are AGN-dominated at optical wavelengths (\S
  \ref{sec:photoz}), while the solid circles are the other sources in
  our sample. The dashed lines correspond to the 1$\sigma$ dispersion
  $\sigma_z=0.14(1 +z)$ measured for the \citet{cohen2000} sample at
  $z<1.1$.  For the AGN-dominated sources, the redshift is based upon
  the best fit to the quasar spectral template discussed in \S
  \ref{sec:photoz}.
    ($b$) Same as $a$, but with a more restricted redshift range
  ($z<1.7$) and with the AGN-dominated sources excluded.
\label{fig:zphot}}
\end{figure*}

Of the 66 sources, only four are in the HDF proper and have previous
published photometric redshifts using 7-color {\it HST} photometry.
For one galaxy, HDF 36569\_1302, \citet{fernandezsoto2001} dispute the
spectroscopic redshift, arguing that the published value is
attributable to the halo of a nearby, brighter galaxy. Due to this
uncertainty, we exclude HDF 36569\_1302 from our analysis, yielding a
final sample of 65 sources.  Figure \ref{fig:scatter} shows the
distribution of optical and X-ray properties in our subsample,
compared to properties of the rest of the extragalactic sources from
\citet{brandt2001} that lie within the CFGRS region. Our subsample
includes 80\% of the extragalactic sources with optical counterparts
brighter than $R=24.5$ (90\% with counterparts brighter than $R=24$).
Since spectroscopic redshifts are lacking for sources with $R>24.5$ we
are unable to study this subset of the population, which includes the
majority of faint X-ray sources with high $f_X/f_R$. We emphasize that
the results presented in this paper are not necessarily applicable to
these sources.

\section{Photometric Redshifts}
\label{sec:photoz}

\subsection{Redshift Estimation Technique}

We utilize the Bayesian photometric redshift code of
\citet{benitez2000}, BPZ, to derive photometric redshifts for our
sample.  This code employs spectral template fitting, but also takes
into account the $I-$band magnitude of the galaxy. Specifically, the
code utilizes the F814 magnitude distribution of the Hubble field as a
Bayesian prior probability distribution, which enables superior
discrimination when color degeneracies exist between low- and
high-redshift galaxies.  For the current analysis we have only four
filters, so the Bayesian magnitude prior is necessary to break these
color degeneracies and minimize the number of spurious outliers.  Our
results should be valid for any code that utilizes spectral template
fitting though, and to verify this expectation we also derive
photometric redshifts using another publicly available code, Hyperz
\citep{bolz2000}.  While BPZ is more robust to outliers for our data
set, we confirm with Hyperz that our results are insensitive to which
photometric code is used.

The standard six spectral templates described in \cite{benitez2000}
are used for this analysis. Four are based upon spectral energy
distributions (SED's) in \citet{cww1980} (E/S0, Sbc, Scd, and Irr),
and two are derived from spectra for starburst galaxies in
\citet{kinney1996}.  One challenge for the current analysis is that we
lack $I$-band data for comparison with the HDF magnitude priors.
Fortunately, the code is not strongly sensitive to errors in the
$I$-band magnitude used for the prior, so we make the coarse
assumption that ${\cal R}-F814\simeq1$ (comparable to an elliptical
galaxy at $z\simeq0.5$) for all galaxies in the sample. We also test
the code with ${\cal R}-F814=0$, finding that the choice of fiducial
color yields a negligible difference in estimated redshifts.

We employ a two-part approach in deriving photometric redshifts. A
necessary first step is to identify objects with quasar-dominated
spectra at optical wavelengths, since photometric redshifts based upon
galaxy spectral templates are likely to fail for these sources. If
these objects can be identified, then photometric redshifts can
potentially be derived for them by other means \citep[e.g.
see][]{budavari2001,richards2001}. To this end, we construct a quasar
template using the optical SDSS composite spectrum of \citet{van2001},
coupled with infrared spectra of PDS 456 from \citet{simpson1999}. For
the SDSS composite, we apply a correction for stellar contamination,
assuming that the contamination increases linearly from 10\% at
6000$\AA$ to 30\% at 8500$\AA$, which is based upon estimates in
\citet{van2001}. The resulting template is shown in Figure
\ref{fig:qsotemplate}.

We initially include this quasar spectral template in the SED library
and run the photometric redshift code without the Bayesian prior.  Six
of the 65 sources in our sample, including the two highest redshift
sources,{\footnote{This is a selection effect -- if a source at
$z\simeq2$ is not a quasar, then it will be optically faint and
likely not have been targeted for spectroscopy. }} are best fit by
the quasar spectrum. One of these sources has previously been
identified as a quasar \citep{liu1999,horn2001}; a second was
identified by \citet{horn2001} as a BL AGN.  We exclude these six
sources from the subsequent analysis, but record the photometric
redshift estimates and include them in subsequent Figures.  For the
remaining 59 X-ray sources, all of which are at $z<1.3$, we perform a
second iteration of the redshift estimation in which we exclude the
quasar spectrum and include the Bayesian magnitude priors.  The
results are shown in Figures \ref{fig:zphot}$a,b$. We derive a scatter
$\sigma_z=0.10(1+z)$ for these sources.

\begin{inlinefigure}
\plotone{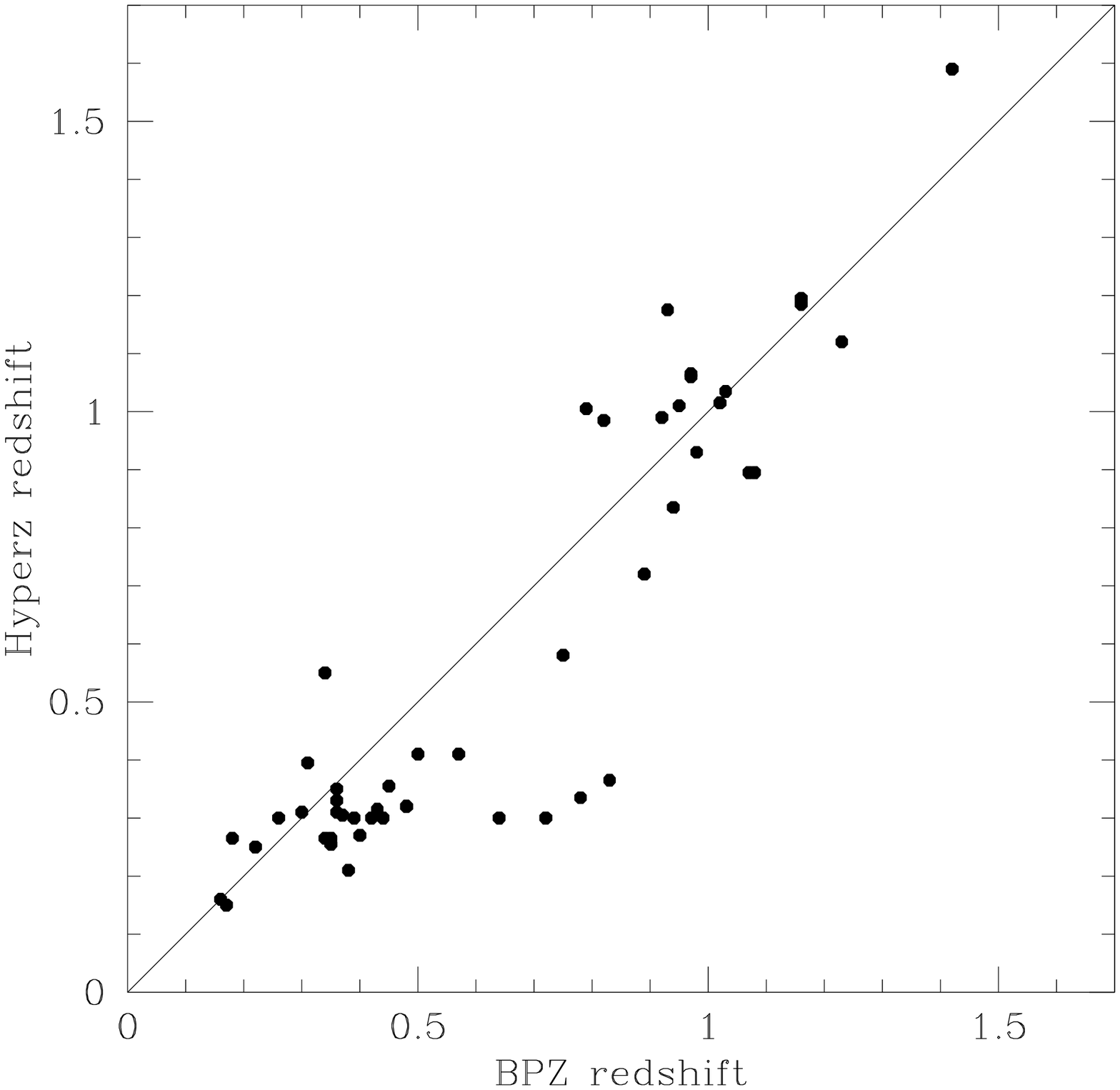} 
\figcaption{
  Comparison of photometric redshifts obtained with Hyperz and BPZ for
  the X-ray luminous sources. We only plot objects that are neither
  quasar-dominated nor outliers with Hyperz (as defined in the text).
\label{fig:hyperz}}
\end{inlinefigure}

\subsection{Comparison with Normal Galaxy Sample}
\label{sec:comparison}
\citet{cohen2000} provide spectroscopic redshifts for a total of 529
galaxies in the \citet{hogg2000} photometric catalog, enabling
construction of an X-ray quiet comparison sample. Excluding the {\it
Chandra} sources and galaxies at $z>1.3$, the redshift dispersion for
this catalog is $\sigma_z=0.18(1+z)$.  This scatter is dominated by a
handful of outliers -- if we employ $\sigma$-clipping with rejection
of outliers beyond 3.5$\sigma$, then 11 out of 461 galaxies are
rejected and the dispersion reduces to $\sigma_z=0.14(1+z)$. This
value is larger than the scatter for the HDF
\citep[$\sigma_z\la0.10(1+z)$ for 7 filters; e.g.][]{fernandezsoto1999,
benitez2000,bolz2000,furusawa2000}, but not unreasonable given the
larger photometric uncertainties and smaller number of passbands in
this analysis.  Comparing with the X-ray subsample, we conclude that,
with the exception of clearly quasar-dominated sources, the dispersion
in photometric redshifts is no greater for X-ray sources than for the
field galaxy population.

To verify these results, we also derive photometric redshifts using
Hyperz. The Hyperz code was utilized by \citet{barger2002}, who
concluded that the lack of priors coupled with their lack of $U-$band
data significantly hindered the reliability of their photo-$z$'s. Our
comparison of BPZ and Hyperz confirms that the inclusion of a
magnitude prior significantly improves the robustness, particularly
with a limited number of passbands. With Hyperz, 12\% of the sources
in the quiescent sample -- and a comparable fraction of the X-ray
luminous sources -- are outliers with $\Delta z>1$. In contrast, only
2\% of sources in the quiescent sample are outliers with BPZ.
Nonetheless, if these outliers are excluded, then Hyperz and BPZ
typically yield consistent photometric redshifts for individual X-ray
selected sources (Figure \ref{fig:hyperz}).\footnote{Hyperz does
  appear to underestimate the redshift for sources at $z\sim0.7$
  though.} Furthermore, Hyperz yields dispersions comparable to BPZ,
with values $\sigma_z=0.14(1+z)$ for the full field sample and
$\sigma_z=0.13(1+z)$ for the X-ray sources.  Both codes thus

\begin{inlinefigure}
\plotone{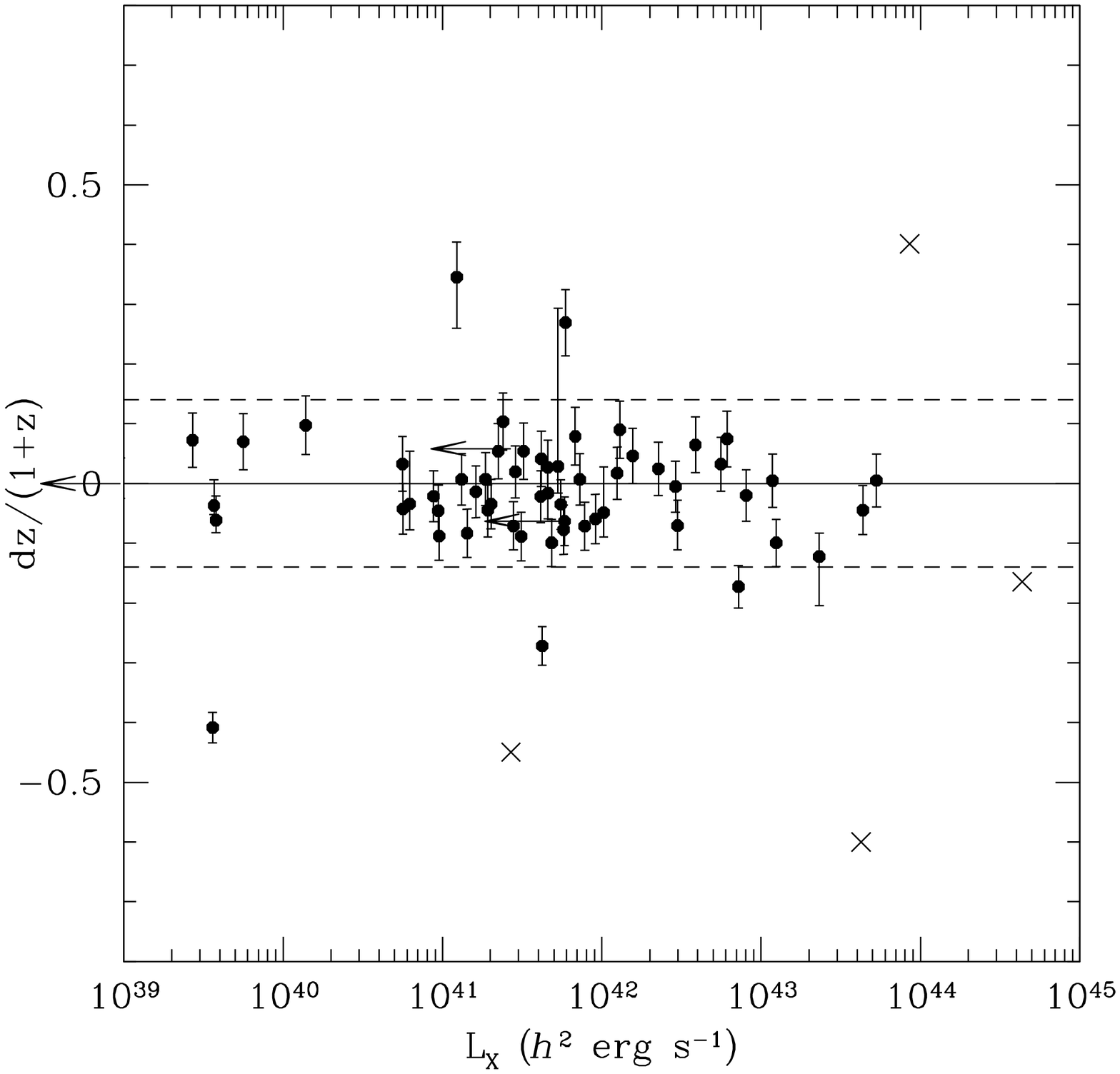} 
\figcaption{
  Photometric redshift error as a function of X-ray luminosity. The
  symbols are the same as in Figure \ref{fig:zphot}. The dashed lines
  correspond to $\sigma_z=0.14(1+z)$. The vertical axis is set to
  match the dynamic range of the non-quasar-dominated sources.
  Consequently, two highly discrepant quasar-dominated sources with
  $dz/(1+z)\approx1.5$ and $L_X\approx4\times10^{42}$ erg s$^{-1}$
  cm$^{-2}$ are not visible in this plot.
\label{fig:luminosity}}
\end{inlinefigure}
\begin{inlinefigure}
\plotone{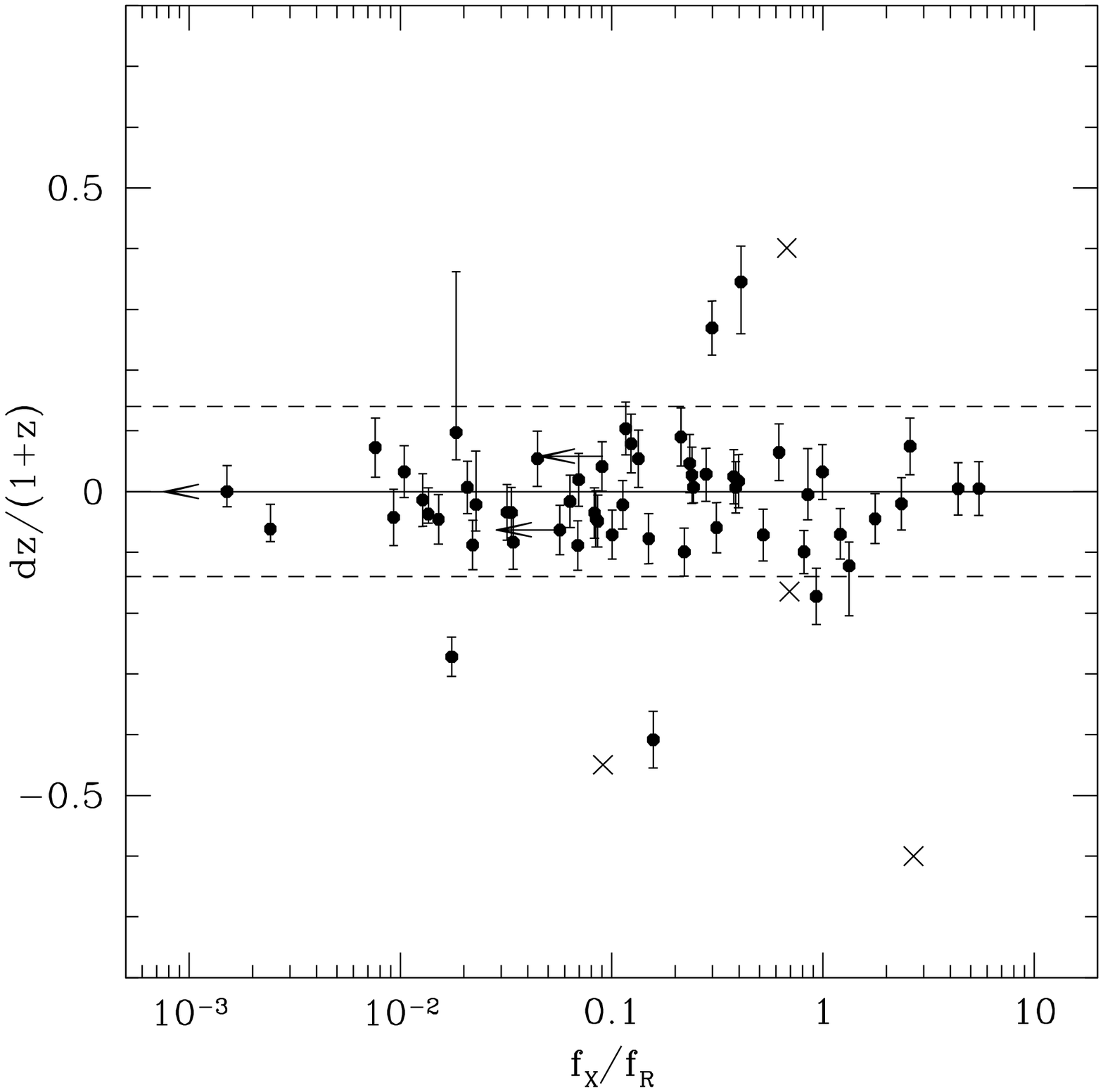} 
\figcaption{
    Photometric redshift error as a function of $f_X/f_R$. The symbols
  are the same as in Figure \ref{fig:zphot}, and the dashed lines
  correspond to $\sigma_z=0.14(1+z)$. As in Figure
  \ref{fig:luminosity}, two quasar-dominated sources with
  $dz/(1+z)\approx1.5$ and $f_X/f_R\approx 2-4$ are not visible in
  this plot.
\label{fig:fxfr}}
\end{inlinefigure}

\noindent indicate
that the dispersion is comparable for the X-ray luminous and quiescent
samples.

\subsection{Dependence upon X-ray Properties}

The results of the preceding sections indicate that photometric
redshifts are robust for $\sim$90\% of X-ray sources with $R\la24.5$
down to $f_X\simeq10^{-16}$ erg s$^{-1}$ cm$^{-2}$, including some
sources with a moderate AGN contribution.  One naively expects
increased scatter in the photometric redshifts with increasing optical
contribution from the AGN. A direct test of this expectation would
require fitting a linear superposition of normal galaxy and quasar
spectral templates to obtain the fractional contribution, which is
beyond the scope of the current analysis. Instead, we ask whether the
probability of failure is correlated with the X-ray properties of the
source -- specifically the X-ray luminosity and the ratio of X-ray to
optical flux ($f_X/f_R$).  The latter quantity is the best analog
measure of the relative AGN and starlight contributions, but with the
caveat that there exists a factor of ten range in $f_X/f_R$ within
which sources can be either AGN or starlight dominated.

In Figure \ref{fig:luminosity} we plot $dz/(1+z)$ against $L_X$, where
$d_z\equiv z_{phot}-z_{spec}$, to assess whether the redshift
estimates degrade above some threshold luminosity.  Aside from objects
that are best fit by the quasar spectral template, there is no
indication of degradation up to $L_X=5\times10^{43} h^2$ erg s$^{-1}$,
the luminosity of the brightest non-quasar-dominated sources in our
sample.  We also find no statistically significant trend with
$f_X/f_R$ (Figure \ref{fig:fxfr}); however, the probability of
identifying a source as quasar-dominated increases significantly for
sources with $f_X/f_R\ga0.5$. Nearly 30\% of sources above this
threshold (5 of 18) are best fit by the quasar template, as opposed to
only one of 47 at lower $f_X/f_R$. These results imply, somewhat
surprisingly that the photometric redshift estimates are fairly
insensitive to the presence of an AGN unless the AGN dominates the
optical spectrum.

\section{Summary}
\label{sec:summary}

We use a sample of X-ray sources detected by \chandra in the Caltech
Faint Galaxy Redshift Survey region to test the robustness of
photometric redshifts for these objects. For the 59 out of 65 sources
with colors that are not best fit by a quasar spectral template, we
find no degradation in the accuracy of photometric redshifts as
compared to the field galaxy population. We also find that the
redshift residuals exhibit no trend as a function of X-ray luminosity
or $f_X/f_R$ for these sources.  We demonstrate that it is feasible to
quickly identify objects whose spectra are quasar-dominated at optical
wavelengths and derive robust photometric redshifts for the other
$\sim90$\% of sources that have optical counterparts brighter that
$R\approx24.5$, which includes roughly two thirds of all sources with
$f_X>10^{-16}$ erg s$^{-1}$ cm$^{-2}$ \citep{barger2002}.
Consequently, photometric redshift estimation should be a valuable
tool for upcoming large samples of optically bright, X-ray selected
objects.  This paper is a first step towards quantifying the
robustness of photometric redshifts for systems in the transition
regime between quiescent galaxies and quasars. The next required step
is extension to fainter optical magnitudes.  Given that the majority
of fainter sources have high $f_X/f_R$, it is unclear whether
photometric redshifts will be as effective for this population, but
the motivation for testing their robustness is strong. There is
speculation that these optically faint, high $f_X/f_R$ sources
possibly include obscured AGN at $z\ga5$ \citep[e.g.][]{stern2002},
and photometric redshifts may be the only viable approach for studying
this population.

\section{Acknowledgements}

The authors gratefully thank Amy Barger for providing a text version
of Table 1 from \citet{barger2002}, Chris Simpson for providing the
infrared spectroscopic data for PDS 456, Mark Dickinson for answering
questions regarding CXOHDFN J123651.7+621221, and Narciso Benitez for
email assistance with BPZ.  Both authors wish to thank the anonymous
referee for the professionalism of the review and for suggestions that
improved the clarity and content of this paper.

\clearpage
\begin{deluxetable}{rrrlllc}
\tablecaption{\chandra Sources with Spectroscopic Redshifts}
\tablewidth{340pt}
\tabletypesize{\footnotesize}
\tablehead{
\colhead {B01\tablenotemark{a}} & 
\colhead {H01\tablenotemark{b}} & 
\colhead {$f_x$\tablenotemark{c}} & 
\colhead {$z$\tablenotemark{d}} & 
\colhead {$z_{phot}$\tablenotemark{e}} & 
\colhead {BL?\tablenotemark{f}} & 
\colhead {Comments\tablenotemark{g}}  
}
\startdata
 080 &    & 0.54    & 0.454 & 0.40 &   &  \\
 087 &    & 0.52    & 0.534 & 0.56 &   &  \\
 089 &  6 & 16.08   & 1.022 & 0.93 & Y &  \\
 092 &    & 0.32    & 0.473 & 0.48 &   &  \\
 098 & 12 & 3.04    & 1.014 & 0.97 &   &  \\
 101 & 14 & 16.33   & 2.590 & Q    & Y & QSO (Liu et al. 1999) \\
 102 &    & 0.22    & 0.483 & 0.35 &   &  \\
 105 &    & 1.00    & 0.747 & 0.78 &   &  \\
 110 &    & $<$0.14 & 1.218 & 1.08 &   &  \\
 111 & 16 & 3.59    & 0.762 & Q    &   &  \\
 117 & 19 & 2.31    & 1.013 & 1.16 &   &  \\
 118 & 20 & 2.27    & 0.848 & 0.97 &   &  \\
 121 &    & 1.53    & 0.953 & Q    &   &  \\
 125 &    & 0.25    & 0.845 & 0.34 &   &  \\
 131 &    & 0.85    & 1.016 & 1.07 &   &  \\
 134 & 26 & 0.43    & 0.456 & 0.44 &   &  \\
 136 &    & 0.31    & 1.219 & 1.42 &   &  \\
 139 &    & 0.19    & 0.847 & 0.95 &   &  \\
 141 & 27 & 0.95    & 0.410 & 0.31 &   &  \\
 142 & 28 & 2.97    & 2.005 & Q    &   &  \\
 144 & 29 & 10.09   & 0.957 & 0.72 &   &  \\
 145 & 30 & 4.94    & 0.555 & 0.45 &   &  \\
 148 &    & $<$0.12 & 0.081 & 0.08 &   &  \\
 149 &    & 0.58    & 0.357 & 0.83 &   &  \\
 154 &    & 0.98    & 0.510 & 0.36 &   &  \\
 155 &    & 0.27    & 0.848 & 0.82 &   &  \\
 156 & 34 & 0.26    & 0.943 & 0.79 &   &  \\
 159 &    & 1.33    & 0.518 & 0.64 &   &  \\
 160 & 36 & 0.38    & 0.089 & 0.05 &   & $z_{FS}=0.10$   \\
 163 & 38 & 3.20    & 0.857 & 0.92 &   &  \\
 165 &    & 0.21    & 1.011 & 0.94 &   &  \\
 171 & 40 & 22.79   & 0.961 & 0.97 &   & $z_{FS}=0.85$   \\
 172 &    & 0.53    & 0.851 & 0.74 &   &  \\
 176 & 43 & 0.75    & 0.475 & 0.34 &   & $z_{FS}=0.38$    \\
 178 &    & 0.15    & 0.139 & 0.07 &   &  \\
 183 &    & 0.09    & 0.089 & 0.17 &   &  \\
 185 &    & 0.15    & 0.475 & 0.42 &   &  \\
 188 &    & 0.30    & 0.410 & 0.38 &   &  \\
 190 & 46 & 2.81    & 0.401 & 0.30 &   &  \\
 193 & 48 & 0.57    & 0.321 & 0.26 &   &  \\
 194 &    & 0.22    & 1.275 & 1.16 &   &  \\
 196 &    & $<$0.14 & 0.890 & Q    &   &  \\
 198 & 49 & 1.27    & 0.955 & 0.94 &   &  \\
 202 &    & 0.11    & 0.517 & 0.57 &   &  \\
 203 & 51 & 0.86    & 0.474 & ---  &   & Excluded; $z_{FS}=1.27$ \\
 205 & 52 & 0.27    & 0.847 & 0.89 &   &  \\
 206 &    & 0.24    & 0.663 & 0.75 &   &  \\
 209 & 53 & 0.23    & 0.137 & 0.22 &   &  \\
 212 & 54 & 7.31    & 0.678 & 0.39 &   &  \\
 220 & 57 & 0.15    & 0.136 & 0.16 &   &  \\            
 223 & 59 & 24.74   & 0.514 & 0.36 & Y &  \\
 225 &    & 0.87    & 0.359 & 0.37 &   &  \\
 228 & 62 & 1.72    & 0.377 & 0.35 &   &  \\
 229 &    & 0.23    & 0.561 & 0.43 &   &  \\
 237 & 65 & 31.94   & 1.018 & Q    & Y &  \\
 239 &    & 0.09    & 1.017 & 1.23 &   &  \\
 244 &    & 0.19    & 0.940 & 1.02 &   &  \\
 245 &    & 0.72    & 0.936 & 1.03 &   &  \\
 246 &    & 0.18    & 0.422 & 0.36 &   &  \\
 247 &    & 0.30    & 0.569 & 0.50 &   &  \\
 264 & 68 & 1.75    & 0.475 & 0.48 &   &  \\
 269 &    & 0.17    & 1.084 & 1.14 &   &  \\
 274 & 72 & 4.40    & 1.019 & 1.03 &   &  \\
 278 &    & 0.18    & 0.230 & 0.35 &   &  \\
 279 &    & 0.97    & 0.557 & 0.98 &   &  \\
 290 &    & 0.20    & 0.104 & 0.18 &   &  \\
\enddata
\tablenotetext{a}{Identification number from \citet{brandt2001}.}
\tablenotetext{b}{Identification number from \citet{horn2001}.}
\tablenotetext{c}{
Full band flux (0.5-8 keV) in units of 10$^{-15}$ erg s$^{-1}$ cm$^{-2}$ from
\citet{brandt2001}.}
\tablenotetext{d}{
Spectroscopic redshifts are taken from \citet{barger2002}. The one exception is source 220,
for which the redshift is taken from \citet{cohen2000}.
}
\tablenotetext{e}{'Q' denotes that the photometry for this source was best fit by
a quasar spectral template.}
\tablenotetext{f}{'Y' denotes that this source was found to have broad emission lines
in the spectra by \citet{barger2002}.}
\tablenotetext{g}{For sources in the HDF, we list the photometric redshift from 
\citet{fernandezsoto2001}. Source 203 was excluded from our analysis due the disputed
redshift.}
\label{tab:tab1}
\end{deluxetable}

\end{document}